\documentclass[12pt]{article}

\usepackage{amstext, graphicx, amsmath, latexsym, amssymb, amscd, amsfonts, float,placeins,color}
\usepackage{multirow}
\usepackage{booktabs}
\usepackage{bbm}
\usepackage{harpoon}
\usepackage{amsthm}
\usepackage{tikz}
\usepackage{siunitx}
\usepackage{amsmath,bm}
\usepackage[title]{appendix}

\usepackage[noend]{algpseudocode}
\usepackage{algorithmicx,algorithm}
\usepackage[colorlinks,linkcolor=blue]{hyperref}
\usepackage{xcolor}

\newtheorem{Theorem}{\bf Theorem}[section]		
\newtheorem{Lemma}{\bf Lemma}[section]
\newtheorem{Proposition}{\bf Proposition}[section]
\newtheorem{Definition}{\bf Definition}[section]
\newtheorem{Example}{\bf Example}[section]
\newtheorem{Remark}{\bf Remark}[section]

\numberwithin{equation}{section}
\numberwithin{figure}{section}
\allowdisplaybreaks[4]

\begin{document}

\title{Dynamic risk measures for fluctuations in market volatility under Bochner–Lebesgue spaces}

\author{Fei Sun$^{1}$, Jingchao Li$^{2}$\footnote{Corresponding Author: Jingchao Li,  E-mail: jingchaoli@szu.edu.cn}, Jieming Zhou$^{3}$\footnote{Corresponding Author: Jieming Zhou,  E-mail:jmzhou@hunnu.edu.cn }
}
\date{}
\maketitle

\begin{center}\small{$^1$ School of Mathematics and Computational Science, Wuyi University, Jiangmen, 529020, P.R. China}
\end{center}
\begin{center} \small {$^2$School of Mathematical Sciences, Shenzhen University, Shenzhen, 518000, P.R.China} 
\end{center}
\begin{center} \small {$^3$MOE-LCSM, School of Mathematics and Statistics, Hunan Normal University, Changsha, 410081, P.R. China} 
\end{center}

\begin{abstract}

\noindent  Starting from the global financial crisis to the more recent disruptions brought about by geopolitical tensions and public health crises, the volatility of risk in financial markets has increased significantly. This underscores the necessity for comprehensive risk measures capable of capturing the complexity and heightened fluctuations in market volatility. This need is crucial not only for new financial assets but also for the traditional financial market in the face of a rapidly changing financial environment and global landscape. In this paper, we consider the risk measures on a special space $L^{p(\cdot)}$, where the variable exponent $p(\cdot)$ is no longer a given real number as in the conventional risk measure space $L^{p}$, but rather a random variable reflecting potential fluctuations in volatility within financial markets. Through further development of axioms related to this class of risk measures, we also establish dual representations for them.

\medskip


\end{abstract}
\thispagestyle{empty}
\newpage

\section{Introduction}

\section{Preliminaries}
\label{sec:2}

\section{Convex risk measures on $\mathbf{L^{p(\cdot)}}$}
\label{sec:3}

~~~~~~Recently, many external factors including changes in international situations, increase of war risk and significant environmental changes, all cause the financial market becomes much more volatile than before, and the financial market also has different orders of risk data mixed over a short period of time. Therefore, it is valuable to study the risk measures on variable exponent Bochner--Lebesgue spaces. Under this position space, the order risk position is no longer a fixed positive number, but a measurable function. The characteristics of this space are able to characterise risk positions in the above volatile financial market.

The main purpose of this paper is to study the dynamic risk measures on variable exponent Bochner--Lebesgue spaces. This section firstly considers convex risk measures on $L^{p(\cdot)}$, which will be used later for the dynamic risk measures. 
In the absence of ambiguity, we denote the variable exponent Bochner--Lebesgue space by $L^{p(\cdot)}:=L^{p(\cdot)}(\Omega, E)$. Let $T$ be a discrete time horizon which can reach infinity and consider a filtered probability space $(\Omega,\mathcal{F},(\mathcal{F}_{t})_{t=0}^{T},\mu)$ with $\{\emptyset,\Omega\}=\mathcal{F}_{0}\subset \mathcal{F}_{1}\subset\ldots \subset\mathcal{F}_{T}=\mathcal{F}$. Let $L^{p(\cdot)}(\mathcal{F}_{t})$ be the space of all strongly $\mathcal{F}_{t}$-measurable functions $f_{t}$ which satisfy Definition~\ref{D22}. Note that $L^{p(\cdot)}=L^{p(\cdot)}(\mathcal{F}_{T})$. We denote 
\[L^{p(\cdot)}(K):=\{f\in L^{p(\cdot)} | f:\Omega\rightarrow K\},\] and \[L^{p'(\cdot)}(K_{0}):=\{g\in L^{p'(\cdot)} | g:\Omega\rightarrow K_{0}\}.\] We then denote the space of all essentially bounded $\mathcal{F}_{t}$-measurable random variables by $L_{t}^{\infty}:=L^{\infty}(\Omega,\mathcal{F}_{t},\mu)$.
The definition of a convex risk measure on $L^{p(\cdot)}$ is introduced by using an axiomatic approach.

\begin{Definition}\label{D31}\textnormal
	{
	A continuous function $\varrho^{K}:L^{p(\cdot)}\rightarrow \mathbb{R}$ is said to be a $p(\cdot)$-convex risk measure (with partial order $ \leq_{K} $) if it satisfies the following conditions.
		\begin{description}
			\item[A1] Monotonicity: for any $f_{1},f_{2}\in L^{p(\cdot)}$, $f_{1}\leq_{K}f_{2}\ a.s. \textrm{ implies }\varrho^{K}(f_{1})\geq \varrho^{K}(f_{2})$;
			\item[A2] Translation invariance: for any $m\in \mathbb{R}$ and $f\in L^{p(\cdot)}$, $\varrho^{K}(f+mz)=\varrho^{K}(f)-m$;
			\item[A3] Convexity: for any $f_{1},f_{2}\in L^{p(\cdot)}$ and $\lambda\in(0,1)$, $\varrho^{K}(\lambda f_{1}+(1-\lambda)f_{2})\leq \lambda \varrho^{K}(f_{1})+(1-\lambda)\varrho^{K}(f_{2})$.
		\end{description}
	}
\end{Definition}

\begin{Remark}\label{R32}\textnormal
	{
		In $\mathbf{A1}$, $f_{1}\leq_{K}f_{2}$ means $f_{1}(\omega)\leq_{K}f_{2}(\omega)$ for any $ \omega\in \Omega $ where the partial order $\leq_{K}  $ is defined in Remark~\ref{R21}.
		The interior point $z$ of $K$ in $\mathbf{A2}$ is considered to be the numeraire asset, which means that $mz\in E$ for any $m\in \mathbb{R}$. Before the dual representation of the $p(\cdot)$-convex risk measures is being studied, the acceptance sets should be defined.
	}
\end{Remark}

To better explain the risk measures on the variable exponent Bochner-Lebesgue space, especially the fact that the moments of random variables will change, we provide the following example.
\begin{Example}\label{E31}
	The variable exponent $p(\cdot)$ in the risk position space can be seen as a response made by investors or market regulators under different market conditions. We consider the space $ L^{p(\cdot)}(\Omega, \mathbb{R}) $ and two market environments $\Omega_{A}, \Omega_{B} \in \mathcal{F}$ with $\mu(\Omega_{A})+\mu(\Omega_{B})=1$. The variable exponent $p(\cdot):\Omega\rightarrow [1,+\infty]$  is  given by
	\[
	p(\cdot)=a\mathcal{X}_{\Omega_{A}} + b\mathcal{X}_{\Omega_{B}},
	\]
	where $\mathcal{X}$ is the indicator function with $ a,b\in[1,+\infty] $. The variable exponent Bochne-Lebesgue space $ L^{p(\cdot)}(\Omega, \mathbb{R}) $ consists of all random variables that either belong to $ L^{a}$ or $L^{b} $. When the market is in environment $\Omega_{A}$, the risk position is characterized by $ L^{a}$, and when the market is in environment $\Omega_{B}$, the risk position is characterized by $ L^{b}$.
\end{Example}

The next example provide an interesting interpretation of risk measures on $ L^{p(\cdot)}$.

\begin{Example}
Under the assumptions of Example~\ref{E31}, we consider two types of situations that an investor may encounter when engaging in a certain risky investment, denoted by $\Omega_{A} $ and $\Omega_{B} $ respectively. Since investors have different risk preferences in different investment environments, they will use different confidence levels $ \alpha_{1} $ and $ \alpha_{2} $ when using $ AV@R $ as a tool for risk quantification. We define a new $ AV@R_{(\cdot)} $ on $ L^{p(\cdot)}(\Omega, \mathbb{R}) $ as follows
\[
AV@R_{(\cdot)} = AV@R_{\alpha_{1}}\mathcal{X}_{\Omega_{A}} + AV@R_{\alpha_{2}}\mathcal{X}_{\Omega_{B}},
\]
	where $\mathcal{X}$ is the indicator function.
\end{Example}

\begin{Definition}\label{D32}\textnormal
	{
		The acceptance set of the $p(\cdot)$-risk measure (with partial order $ \leq_{K} $) $\varrho^{K}$ is defined as
		\begin{displaymath}
			\mathcal{A}_{\varrho^{K}}:=\big\{f\in L^{p(\cdot)}  | \varrho^{K}(f)\leq0\big\}
		\end{displaymath}
		and we denote $\mathcal{A}_{\varrho^{K}}^{0}$ by
		\begin{displaymath}
			\mathcal{A}^{0}_{\varrho^{K}}:=\Big\{g\in \big(L^{p(\cdot)}\big)^{\ast}  | \langle g,f\rangle\geq 0 \textrm{ for any }  f \in \mathcal{A}_\varrho^{K}\Big\}.
		\end{displaymath}
	}
\end{Definition}

\begin{Remark}\label{R33}\textnormal
	{
		It is relatively easy to check that $\mathcal{A}_{\varrho^{K}}$ is a convex set if $\varrho^{K}$ satisfies the convexity property. $\mathcal{A}^{0}_{\varrho^{K}}$ can be considered as the positive polar cone of $\mathcal{A}_{\varrho^{K}}$.
	}
\end{Remark}

Now, the dual representation of $p(\cdot)$-convex risk measures is provided and which will be used in the proof of $p(\cdot)$-dynamic risk measures in Sect.~\ref{sec:5}.

\begin{Theorem}\label{T32}\textnormal
	{
		If $\varrho^{K}:L^{p(\cdot)}\rightarrow \mathbb{R}$ is a $p(\cdot)$-convex risk measure (with partial order $ \leq_{K} $), then for any $f\in L^{p(\cdot)}$,
		\begin{displaymath}
			\varrho^{K}(f)=\sup_{g\in Q_{p(\cdot)}}\{ \langle g,-f\rangle-\alpha(g)\}
		\end{displaymath}
		where
		\begin{displaymath}
			Q_{p(\cdot)}:=\Big\{g\in \big(L^{p(\cdot)}\big)^{\ast}  | \int_{\Omega}\langle \frac{dg}{d\mu}, z\rangle d\mu=1, \frac{dg}{d\mu}\in L^{p'(\cdot)}(K_{0})\Big\},
		\end{displaymath}
		and $\alpha: Q_{p(\cdot)} \rightarrow \mathbb{R}$ is the penalty function while the minimal penalty function $\alpha_{\min}$ is denoted by
		\begin{displaymath}
			\alpha_{\min}(g):=\sup_{f\in L^{p(\cdot)}}\big\{\langle g,-f\rangle-\varrho^{K}(f)\big\}=\sup_{f\in\mathcal{A}_{\varrho^{K}}}\big\{\langle g,-f\rangle\big\}.
		\end{displaymath}
	}
\end{Theorem}

\noindent \textbf{Proof.}
For any $g\in Q_{p(\cdot)}$, denoting
\begin{displaymath}
	\alpha(g)=\sup_{f\in L^{p(\cdot)}}\big\{\langle g,-f\rangle-\varrho^{K}(f)\big\}
\end{displaymath}
and
\begin{displaymath}
	\alpha_{\min}(g)=\sup_{f\in\mathcal{A}_{\varrho^{K}}}\big\{\langle g,-f\rangle\big\}.
\end{displaymath}
We now show that $\alpha(g)=\alpha_{\min}(g)$ for any $g\in Q_{p(\cdot)}$.  For any $f\in\mathcal{A}_{\varrho^{K}}$, $\langle g,-f\rangle-\varrho^{K}(f)\geq\langle g,-f\rangle$. Hence,
\begin{displaymath}
	\sup_{f\in L^{p(\cdot)}}\big\{\langle g,-f\rangle-\varrho^{K}(f)\big\}\geq \sup_{f\in\mathcal{A}_{\varrho^{K}}}\big\{\langle g,-f\rangle-\varrho^{K}(f)\big\}\geq \sup_{f\in\mathcal{A}_{\varrho^{K}}}\big\{\langle g,-f\rangle\big\}.
\end{displaymath}
That is $\alpha(g)\geq \alpha_{\min}(g)$. Then for any $f\in L^{p(\cdot)}$, consider $f_{1}=f+\varrho^{K}(f)z\in \mathcal{A}_{\varrho^{K}}$. Thus,
\begin{eqnarray*}
	\alpha_{\min}(g)\geq \langle g,-f_{1}\rangle &=& \langle g,-f\rangle-\varrho^{K}(f)\langle g,z\rangle\\
	&=& \langle g,-f\rangle-\varrho^{K}(f)\int_{\Omega}\langle \frac{dg}{d\mu}, z\rangle d\mu\\
	&=& \langle g,-f\rangle-\varrho^{K}(f).
\end{eqnarray*}
That is $\alpha(g)\leq \alpha_{\min}(g)$. As $\alpha(g)\geq \alpha_{\min}(g)$ and  $\alpha(g)\leq \alpha_{\min}(g)$, hence we have $\alpha(g)=\alpha_{\min}(g)$, and it is easy to check that
\begin{displaymath}
	\varrho^{K}(f)\geq\sup_{g\in Q_{p(\cdot)}}\big\{ \langle g,-f\rangle-\alpha(g)\big\}.
\end{displaymath}
Next, we show that the above inequality only holds in the case of equality. Suppose there is some $f_{0}\in  L^{p(\cdot)}$ such that
\begin{displaymath}
	\varrho^{K}(f_{0})>\sup_{g\in Q_{p(\cdot)}}\big\{ \langle g,-f_{0}\rangle-\alpha(g)\big\}.
\end{displaymath}
Hence, there exists some $m\in\mathbb{R}$ such that
\begin{displaymath}
	\varrho^{K}(f_{0})>m>\sup_{g\in Q_{p(\cdot)}}\big\{ \langle g,-f_{0}\rangle-\alpha(g)\big\}.
\end{displaymath}
Thus, we have
\begin{displaymath}
	\varrho^{K}(f_{0}+mz)=\varrho^{K}(f_{0})-m>0,
\end{displaymath}
which means that $f_{0}+mz\notin \mathcal{A}_{\varrho^{K}}$. As $\{f_{0}+mz\}$ is a singleton set, it is also a convex set. Meanwhile, $\mathcal{A}_{\varrho^{K}}$ is also a closed convex set because $\varrho^{K}$ is a $p(\cdot)$-convex risk measure. Then, by the Strong Separation Theorem for convex sets, there exists some $\pi\in (L^{p(\cdot)})^{\ast}$ such that
\begin{equation}\label{32}
	\langle \pi,f_{0}+mz\rangle > \sup_{f\in\mathcal{A}_{\varrho^{K}}}\langle \pi,f\rangle.
\end{equation}
By Remark~\ref{R24}, $\langle \pi,f\rangle=\int_{\Omega}\langle h, f\rangle d\mu$, where $h\in L^{p'(\cdot)}(\Omega, E^{\ast})$.  It is easy to check that $\langle \pi,f\rangle=\int_{\Omega}\langle h, f\rangle d\mu \leq 0$ for any  $f\in L^{p(\cdot)}(K)$. Then, we have that $-h\in L^{p'(\cdot)}(K_{0})$. For any $-\pi\in Q_{p(\cdot)}$, $\langle h, z\rangle=-1$. Thus, by (\ref{32}), it gives
\begin{eqnarray*}
	\langle \pi,f_{0}+mz\rangle > \sup_{f\in\mathcal{A}_{\varrho^{K}}}\langle \pi,f\rangle &\Rightarrow& \int_{\Omega}\langle h, f_{0}+mz\rangle d\mu > \sup_{f\in\mathcal{A}_{\varrho^{K}}}\int_{\Omega}\langle h, f\rangle d\mu\\
	&\Rightarrow&\int_{\Omega}\big(\langle h, f_{0}\rangle-m\big) d\mu > \sup_{f\in\mathcal{A}_{\varrho^{K}}}\int_{\Omega}\langle h, f\rangle d\mu\\
	&\Rightarrow&\int_{\Omega}\langle h, f_{0}\rangle d\mu -m> \sup_{f\in\mathcal{A}_{\varrho^{K}}}\int_{\Omega}\langle h, f\rangle d\mu\\
	&\Rightarrow& \langle \pi, f_{0}\rangle-\sup_{f\in\mathcal{A}_{\varrho^{K}}}\langle \pi,f\rangle>m\\
	&\Rightarrow& \langle \pi, f_{0}\rangle-\alpha(-\pi)>m.
\end{eqnarray*}
By replacing $-\pi$ by $g_{0}$, we have
\begin{displaymath}
	\langle g_{0}, -f_{0}\rangle-\alpha(g_{0})>m.
\end{displaymath}
This is a contradiction, because in this case
\begin{displaymath}
	m>\sup_{g\in Q_{p(\cdot)}}\big\{ \langle g,-f_{0}\rangle-\alpha(g)\big\}\geq\langle g_{0}, -f_{0}\rangle-\alpha(g_{0})>m.
\end{displaymath}
The contradiction arises from the assumption that some $f_{0}\in  L^{p(\cdot)}$ exists such that
\begin{displaymath}
	\varrho^{K}(f_{0})>\sup_{g\in Q_{p(\cdot)}}\big\{ \langle g,-f_{0}\rangle-\alpha(g)\big\}.
\end{displaymath}
Hence, we have
\begin{displaymath}
	\varrho^{K}(f)=\sup_{g\in Q_{p(\cdot)}}\big\{ \langle g,-f\rangle-\alpha(g)\big\}.
\end{displaymath}
For the opposite direction, it is relatively simple to check that $\varrho^{K}$ satisfies the properties of a $p(\cdot)$-convex risk measure.
This completes the proof of Theorem~\ref{T32}.\\

\begin{Remark}\textnormal
	{
The $p(\cdot)$-convex risk measures are extensions of convex risk measures in F\"{o}llmer and Schied (2002).
In fact, if the position space $ L^{p(\cdot)} $ degenerates into the ordinary $ L^{p} $ space, then the dual representation in Theorem~\ref{T32} also becomes the ordinary dual representation of convex risk measures.
}
\end{Remark}

An example of the degenerated dual representation of  $p(\cdot)$-convex risk measure in Theorem~\ref{T32} on $ L^{1} $ is given in below. However, before deriving the degenerated dual representation of  $p(\cdot)$-convex risk measure on $ L^{1} $, a cone $C_{0}$ in  $ L^{\infty} $ needs to be found so that the numeraire asset $e $ is an interior point of $C$, where $C$ is the positive dual cone of $C_{0}$. In fact, this cone $C$ can be seen as the degeneration of  $K $ for $L^{p(\cdot)}$.  According to Proposition 2.4 and what is mentioned in Jameson (1970), we have that $e $ is an interior point of $C$ by considering the  cone $ C_{0}=\{ h\in L^{\infty}(\Omega,\mathcal{F},\mu)  | \int_{\Omega}hd\mu \geq\frac{1}{2} \|h\|_{\infty} \} $. 

\begin{Example}\textnormal
	{
Let $L^{p(\cdot)}$ degenerates into $L^{1}(\Omega,\mathcal{F},\mu)$, $ z $ degenerates into	$ e $ and $L^{\infty}(\Omega,\mathcal{F},\mu)$ is partially ordered by the cone $C_{0}=\{ h\in L^{\infty}(\Omega,\mathcal{F},\mu)  | \int_{\Omega}hd\mu \geq\frac{1}{2} \|h\|_{\infty} \}$, then $ e\in int(C) $ and we may suppose that $L^{\infty}(\Omega,\mathcal{F},\mu)$ is partially ordered by the wedge $ C $. 
Then by Feinstein and Rudloff (2013) and Feinstein and Rudloff (2015), we have the risk measures $\varrho^{K}_{deg}: L^{1}(\Omega,\mathcal{F},\mu)\rightarrow \mathbb{R}$ represented in the way that Theorem~\ref{T32} indicates,
\begin{displaymath}
\varrho^{K}_{deg}(X)=\sup_{\mathfrak{Y}\in \mathcal{H}}\{ \mathfrak{Y}(-X)-\beta(\mathfrak{Y}) \},	
\end{displaymath}
where
\begin{displaymath}
\mathcal{H}=\{\mathfrak{Y}\in C_{0} | \mathfrak{Y}(e)=1  \}.
\end{displaymath}
}
\end{Example}

A special example of $p(\cdot)$-convex risk measures which is so called OCE, is discussed in the next section. Finally, in Sect.~\ref{sec:5}, the $p(\cdot)$-convex risk measures are used to study the dual representation of the $p(\cdot)$-dynamic risk measures.

\section{Optimized Certainty Equivalent on $\mathbf{L^{p(\cdot)}}$}
\label{sec:4}
~~~~~~In this section, a special class of $p(\cdot)$-convex risk measures that is the Optimized Certainty Equivalent (OCE) is studied and it will be used as an example of dynamic risk measures in Sect.~\ref{sec:5}.
The OCE was first introduced by Ben-Tal and Teboulle (1986) and later developed by the same researchers Ben-Tal and Teboulle (2007). In this section,  the OCE on variable exponent Bochner--Lebesgue spaces  $L^{p(\cdot)}$ is defined.  Further, we establish its main properties and show that how it can be used to generate $p(\cdot)$-convex risk measures. Note that the OCE can be used as an application of convex risk measures. Other applications of risk measures such as the application in insurance, see Wang and Peng (2017).

\begin{Definition}\label{D41}\textnormal
	{
		Let $u:E\rightarrow [-\infty,+\infty]$ be a closed, concave, and non-decreasing (partially ordered by $K$) function. Suppose that $u(\theta)=0$, where $\theta$ is the zero element of $E$. We denote the set of such $u$ by $U$.
	}
\end{Definition}

\begin{Remark}\label{R41}\textnormal
	{
		For any $u\in U$ and $f\in L^{p(\cdot)}$, we denote by $u(f):\Omega\rightarrow\mathbb{R}$ and $\mathbb{E}u(f)$ the expectation of $u(f)$ with respect to a probability measure $\mu$.
	}
\end{Remark}

\begin{Definition}\label{D42}\textnormal
	{
		For any $u\in U$ and $f\in L^{p(\cdot)}$, the OCE of some uncertain outcome $f$ is defined by the map $S_{u}:L^{p(\cdot)}\rightarrow\mathbb{R}$,
		\begin{displaymath}
			S_{u}(f)=\sup_{\eta\in \mathbb{R}}\{\eta+\mathbb{E}u(f-\eta z)\}
		\end{displaymath}
		where the domain of $S_{u}$ is defined as dom$S_{u}=\{f\in L^{p(\cdot)}| S_{u}(f)>-\infty\}\neq \emptyset$ and $S_{u}$ is finite on dom$S_{u}$.
	}
\end{Definition}

\begin{Theorem}\label{T41}\textnormal
	{
		For any $u\in U$, the following properties hold for $S_{u}$:
		\begin{description}
			\item[(a)] For any $f\in L^{p(\cdot)}$ and $m\in \mathbb{R}$, $S_{u}(f+mz)=S_{u}(f)+m$;
			\item[(b)] For any $f_{1},f_{2}\in L^{p(\cdot)}$, $f_{1}\leq_{K} f_{2}$ a.s. implies that $S_{u}(f_{1})\leq S_{u}(f_{2})$;
			\item[(c)] For any $f_{1},f_{2}\in L^{p(\cdot)}$ and $\lambda\in (0,1)$, $S_{u}(\lambda f_{1}+(1-\lambda)f_{2})\geq \lambda S_{u}(f_{1})+(1-\lambda)S_{u}(f_{2})$.
		\end{description}
	}
\end{Theorem}

\noindent \textbf{Proof.}
\begin{description}
	\item[(a)] For any $f\in L^{p(\cdot)}$, $m\in \mathbb{R}$,
	\begin{eqnarray*}
		S_{u}(f+mz)&=&\sup_{\eta\in \mathbb{R}}\{\eta+\mathbb{E}u(f+mz-\eta z)\}\\
		&=&m+\sup_{\eta\in \mathbb{R}}\{\eta-m+\mathbb{E}u(f-(\eta-m)z)\}\\
		&=&m+S_{u}(f).
	\end{eqnarray*}
	\item[(b)] For any $f_{1},f_{2}\in L^{p(\cdot)}$ with $f_{1}\leq_{K} f_{2}$, we have $f_{1}-\eta z\leq_{K}f_{2}-\eta z$. As $u$ is non-decreasing, we have
	\begin{displaymath}
		S_{u}(f_{1})=\sup_{\eta\in \mathbb{R}}\{\eta+\mathbb{E}u(f_{1}-\eta z)\}\leq \sup_{\eta\in \mathbb{R}}\{\eta+\mathbb{E}u(f_{2}-\eta z)\}=S_{u}(f_{2}).
	\end{displaymath}
	\item[(c)] For any $f_{1},f_{2}\in L^{p(\cdot)}$ and $\lambda\in (0,1)$,
	\begin{displaymath}
		S_{u}(\lambda f_{1}+(1-\lambda)f_{2})=\sup_{\eta\in \mathbb{R}}\Big\{\eta+\mathbb{E}u\big(\lambda f_{1}+(1-\lambda)f_{2}-\eta z\big)\Big\}.
	\end{displaymath}
	We take $\eta=\lambda\eta_{1}+(1-\lambda)\eta_{2}$. Then,
	\begin{eqnarray*}
		&&S_{u}(\lambda f_{1}+(1-\lambda)f_{2})\\
		&=&\sup_{\eta_{1},\eta_{2}\in \mathbb{R}}\Big\{\lambda\eta_{1}+(1-\lambda)\eta_{2}+\mathbb{E}u\big(\lambda (f_{1}-\eta_{1}z)+(1-\lambda)(f_{2}-\eta_{2}z)\big)\Big\}\\
		&\geq&\sup_{\eta_{1},\eta_{2}\in \mathbb{R}}\Big\{\lambda\eta_{1}+(1-\lambda)\eta_{2}+\lambda \mathbb{E}u (f_{1}-\eta_{1}z)+(1-\lambda)\mathbb{E}u(f_{2}-\eta_{2}z)\Big\}\\
		&=&\sup_{\eta_{1},\eta_{2}\in \mathbb{R}}\Big\{\lambda\big(\eta_{1}+\mathbb{E}u (f_{1}-\eta_{1}z)\big)+(1-\lambda)\big(\eta_{2}+\mathbb{E}u (f_{2}-\eta_{2}z)\big)\Big\}\\
		&=&\lambda S_{u}(f_{1})+(1-\lambda)S_{u}(f_{2}).
	\end{eqnarray*}
\end{description}
This completes the proof of Theorem~\ref{T41}.\\

\begin{Theorem}\label{T42}\textnormal
	{
		The function $\varrho^{K}$, defined as $\varrho^{K}(f):=-S_{u}(f)$ for any $f\in L^{p(\cdot)}$, is a $p(\cdot)$-convex risk measure.
	}
\end{Theorem}

\noindent \textbf{Proof:}
The proof of Theorem~\ref{T42} is straightforward from Theorem~\ref{T41}.\\

\begin{Proposition}\label{P41}\textnormal
	{
		For any $u\in U$, $\alpha\in \mathbb{R}^{+}$, and $f\in L^{p(\cdot)}$, the OCE $S_{u}(f)$ is sub-homogeneous, i.e.
		\begin{description}
			\item[(a)] $S_{u}(\alpha f)\leq \alpha S_{u}(f), \qquad \forall\alpha>1$;
			\item[(b)] $S_{u}(\alpha f)\geq \alpha S_{u}(f), \qquad \forall0\leq\alpha\leq1$.
		\end{description}
	}
\end{Proposition}

\noindent \textbf{Proof.}
Denote $S(\alpha):=\frac{1}{\alpha}S_{u}(\alpha f)$, then,
\begin{equation}\label{41}
	S(\alpha)=\frac{1}{\alpha}S_{u}(\alpha f)=\sup_{\eta\in \mathbb{R}}\Big\{\eta+\mathbb{E}\frac{1}{\alpha}u\big(\alpha(f-\eta z)\big)\Big\}.
\end{equation}
Next, we show that $S(\alpha)$ is non-increasing in $\alpha>0$ for any $f\in L^{p(\cdot)}$. For $\alpha_{2}\geq\alpha_{1}\geq0$, we have
\begin{displaymath}
	\frac{u(\alpha_{2}t)-u(\alpha_{1}t)}{\alpha_{2}-\alpha_{1}}\leq \frac{u(\alpha_{1}t)-u(\theta)}{\alpha_{1}-0},\qquad\textrm{for any $t\in E$}
\end{displaymath}
by the concavity of $u$. As $u(\theta)=0$, we have
\begin{displaymath}
	\frac{1}{\alpha_{2}}u(\alpha_{2}t)\leq \frac{1}{\alpha_{1}}u(\alpha_{1}t).
\end{displaymath}
Then, from (\ref{41}), we have $S(\alpha_{1})\geq S(\alpha_{2})$, which clearly implies $\mathbf{(a)}$ and $\mathbf{(b)}$.\\

\begin{Proposition}\label{P42}\textnormal
	{
		(Second-order stochastic dominance)
		We denote $C_{u}(f)$ by $uC_{u}(f):=\mathbb{E}u(f)$ for any $u\in U$ and $f\in L^{p(\cdot)}$. We also assume that the supremum in the definition of $S_{u}$ is attained. Then, for any $f_{1}, f_{2}\in L^{p(\cdot)}$,
		\begin{displaymath}
			S_{u}(f_{1})\geq S_{u}(f_{2})\qquad \textrm{if and only if}\qquad C_{u}(f_{1})\geq C_{u}(f_{2}).
		\end{displaymath}
	}
\end{Proposition}

\noindent \textbf{Proof.}
At first, we show the ``if'' part. If $C_{u}(f_{1})\geq C_{u}(f_{2})$, we have $\mathbb{E}{u}(f_{1})\geq \mathbb{E}{u}(f_{2})$ by the fact that $u$ is non-decreasing.
From the definition of $S_{u}$, it follows that $S_{u}(f_{1})\geq S_{u}(f_{2})$. Then we show the ``only if'' part. Let $\ell_{f_{1}},\ell_{f_{1}}$ be the points where the suprema of $S_{u}(f_{1})$ and
$S_{u}(f_{2})$ are attained, respectively. Then, for any $u\in U$,
\begin{displaymath}
	S_{u}(f_{1})=\ell_{f_{1}}+\mathbb{E}u(f_{1}-\ell_{f_{1}} z)\geq \ell_{f_{2}}+\mathbb{E}u(f_{2}-\ell_{f_{2}} z)\geq \ell_{f_{1}}+\mathbb{E}u(f_{2}-\ell_{f_{1}} z),
\end{displaymath}
where the first inequality comes from $S_{u}(f_{1})\geq S_{u}(f_{2})$. Therefore, for any $u\in U$,\\ $\mathbb{E}u(f_{1}-\ell_{f_{1}} z)\geq \mathbb{E}u(f_{2}-\ell_{f_{1}} z)$, which implies $\mathbb{E}u(f_{1})\geq \mathbb{E}u(f_{2})$. Then, $C_{u}(f_{1})\geq C_{u}(f_{2})$.\\

\section{Dynamic risk measures on $\mathbf{L^{p(\cdot)}}$}
\label{sec:5}

In this section, we extend the definition of $p(\cdot)$-convex risk measures in Sect.~\ref{sec:3} to a dynamic setting.
Similar to the axiomatic approach in Detlefsen and Scandolo (2005), we first define the
conditional $p(\cdot)$-convex risk measures.

\begin{Definition}\label{D51}\textnormal
	{
		A map $\varrho^{K}_{t}:L^{p(\cdot)}$ $\rightarrow$ $ L_{t}^{\infty}$ is called a conditional $p(\cdot)$-convex risk measure (with partial order $ \leq_{K} $) if it satisfies the following properties for all $f, f_{1}, f_{2}\in L^{p(\cdot)}$:
		\begin{description}
			\item[i.] Monotonicity: $f_{1}\leq_{K}f_{2}\ a.s. \textrm{ implies }\varrho^{K}_{t}(f_{1})\geq \varrho^{K}_{t}(f_{2})$;
			\item[ii.] Conditional cash invariance: for any $m_{t}\in L_{t}^{\infty}$, $\varrho^{K}_{t}(f+m_{t}z)=\varrho^{K}_{t}(f)-m_{t}$;
			\item[iii.] Conditional convexity: for any $\lambda\in L_{t}^{\infty}$ with $\lambda\in[0,1]$, $\varrho^{K}_{t}(\lambda f_{1}+(1-\lambda)f_{2})\leq \lambda \varrho^{K}_{t}(f_{1})+(1-\lambda) \varrho^{K}_{t}(f_{2})$;
			\item[iv.] Normalization: $\varrho^{K}_{t}(\theta)=0$, $\varrho^{K}_{t}(f)<\infty$.
		\end{description}
	}
\end{Definition}

\begin{Remark}\label{R51}\textnormal
	{
		Note that any element in $L_{t}^{\infty}:=L^{\infty}(\Omega,\mathcal{F}_{t},\mu)$ is a random variable, where $\mathcal{F}_{t}$ is a sub-$\sigma$-algebra of $\mathcal{F}$. As stated by Detlefsen and Scandolo \cite{9}, if the additional information is described by a sub-$\sigma$-algebra $\mathcal{F}_{t}$ of the total information $\mathcal{F}_{T}$, then a conditional risk measure is a map assigning an $\mathcal{F}_{t}$-measurable random variable $\varrho^{K}_{t}(f)$, representing the conditional riskiness of $f$, to every $\mathcal{F}_{T}$-measurable function $f$, representing a final payoff.
	}
\end{Remark}

The acceptance set of a conditional $p(\cdot)$-convex risk measure $\varrho^{K}_{t}$ is defined as
\begin{equation}\label{51}
	\mathcal{A}_{t}:=\big\{f\in L^{p(\cdot)}  | \varrho^{K}_{t}(f)\leq0\big\} \textrm{ for any } 0\leq t\leq T.
\end{equation}
The corresponding stepped acceptance set is defined as
\begin{equation}\label{52}
	\mathcal{A}_{t,t+s}:=\big\{f\in L^{p(\cdot)}(\mathcal{F}_{t+s})  | \varrho^{K}_{t}(f)\leq0\big\} \textrm{ for any } 0\leq t<t+s\leq T.
\end{equation}

\begin{Proposition}\label{P51}\textnormal
	{
		The acceptance set $\mathcal{A}_{t}$ of a conditional $p(\cdot)$-convex risk measure $\varrho^{K}_{t}$ has the following properties:
		\begin{description}
			\item[1.] Conditional convexity: for any $f_{1}, f_{2}\in \mathcal{A}_{t}$, and an $\mathcal{F}_{t}$-measurable function $\alpha$ with $0\leq\alpha\leq1$, we have $\alpha f_{1}+(1-\alpha)f_{2}\in \mathcal{A}_{t}$;
			\item[2.] Solidity: for any $f_{1}\in \mathcal{A}_{t}$ with $f_{1}\leq_{K}f_{2}$ a.s. implies $f_{2}\in \mathcal{A}_{t}$;
			\item[3.] Normalization: $\theta\in \mathcal{A}_{t}$.
		\end{description}
	}
\end{Proposition}
\noindent \textbf{Proof.}
It is easy to check properties 1--3 by using Definition~\ref{D51}.\\

\begin{Definition}\label{D52}\textnormal
	{
		A sequence $(\varrho^{K}_{t})_{t=0}^{T}$ is called a dynamic $p(\cdot)$-convex risk measure if each $\varrho^{K}_{t}$ is a conditional $p(\cdot)$-convex risk measure for any $0\leq t\leq T$.
	}
\end{Definition}

We now study the dual representation of a conditional $p(\cdot)$-convex risk measure. At first, the notion of the $\mathcal{F}_{t}$-conditional inner product related to $L^{p(\cdot)}$ should be defined.\\

\begin{Definition}\label{D53}\textnormal
	{
		For any $f\in L^{p(\cdot)}$ and $g\in (L^{p(\cdot)})^{\ast}$, we define the $\mathcal{F}_{t}$-conditional inner product $\langle g,-f\rangle_{t}$ by
		\begin{equation}\label{53}
			\langle g,-f\rangle_{t} = \int_{\Omega}\langle V_{g},-f\rangle d P,
		\end{equation}
	where $V_{g}\in L^{p^{'}(\cdot)}(\Omega, E^{\ast})$ and $ P\in\{P\in(\Omega,\mathcal{F}_{t}) | P \textrm{ is absolutely continuous w.r.t.} \mu, P=\mu \textrm{ on } \mathcal{F}_{t}\} $.
		We also define the  minimal penalty function $\alpha_{t}^{\min}$  as
		\begin{equation}\label{54}
			\alpha_{t}^{\min}(g):=\textrm{ess}\sup_{f\in\mathcal{A}_{t}}\langle g,-f\rangle_{t}.
		\end{equation}
	}
\end{Definition}

\begin{Lemma}\label{L51}\textnormal
	{
		For any $g\in Q_{p(\cdot)}$ and $0\leq t\leq T$,
		\begin{equation}\label{55}
			\int_{\Omega}\alpha_{t}^{\min}(g) dP=\sup_{f\in\mathcal{A}_{t}}\langle g,-f\rangle.
		\end{equation}
	}
\end{Lemma}
\noindent \textbf{Proof.}
We first show that there exists a sequence $(f_{n})_{n\in \mathbb{N}}$ in $\mathcal{A}_{t}$ such that
\begin{equation}\label{56}
	\textrm{ess}\sup_{f\in\mathcal{A}_{t}}\langle g,-f\rangle_{t}=\lim_{n\rightarrow\infty}\langle g,-f_{n}\rangle_{t}.
\end{equation}
Indeed, for any $f_{1}, f_{2}\in \mathcal{A}_{t}$, we define $\widehat{f}:=f_{1}I_{B}+f_{2}I_{B^{c}}$ where $B:=\{\langle g,-f_{1}\rangle_{t}\geq\langle g,-f_{2}\rangle_{t}\}$. By property 1 of Proposition~\ref{P51}, we know that $\widehat{f}\in\mathcal{A}_{t}$. Hence, by the definition of $\widehat{f}$,
\begin{displaymath}
	\langle g,-\widehat{f}\rangle_{t}=\max\{\langle g,-f_{1}\rangle_{t}, \langle g,-f_{2}\rangle_{t}\}.
\end{displaymath}
Thus, (\ref{56}) holds. We now have
\begin{eqnarray*}
	\int_{\Omega}\alpha_{t}^{\min}(g) dP&=&\int_{\Omega}\textrm{ess}\sup_{f\in\mathcal{A}_{t}}\langle g,-f\rangle_{t} dP\\
	&=&\int_{\Omega}\lim_{n\rightarrow\infty}\langle g,-f_{n}\rangle_{t} dP\\&=&\lim_{n\rightarrow\infty}\int_{\Omega}\langle g,-f_{n}\rangle_{t} dP\\&=&\lim_{n\rightarrow\infty}\langle g,-f_{n}\rangle\\&\leq&\sup_{f\in\mathcal{A}_{t}}\langle g,-f\rangle.
\end{eqnarray*}
The converse inequality can be checked easily.\\

The following theorem gives the dual representation of conditional $p(\cdot)$-convex risk measures.

\begin{Theorem}\label{T51}\textnormal
	{
		Suppose $\varrho^{K}_{t}$ is a conditional $p(\cdot)$-convex risk measure (with partial order $ \leq_{K} $). Then, the following statements are equivalent.
		\begin{description}
			\item[(1)] $\varrho^{K}_{t}$ has the robust representation
			\begin{equation}\label{57}
				\varrho^{K}_{t}(f)=\textrm{ess}\sup_{g\in Q_{p(\cdot)}}\big\{\langle g,-f\rangle_{t}-\alpha_{t}(g)\big\}\quad \textrm{for any}\quad f\in L^{p(\cdot)},
			\end{equation}
			where
			\begin{displaymath}
				Q_{p(\cdot)}:=\Big\{g\in \big(L^{p(\cdot)}\big)^{\ast}  | \int_{\Omega}\langle \frac{dg}{d\mu}, z\rangle d\mu=1, \frac{dg}{d\mu}\in L^{p'(\cdot)}(K_{0})\Big\},
			\end{displaymath}
			and $\alpha_{t}$ is the penalty function from $Q_{p(\cdot)}$ to the set of $\mathcal{F}_{t}$-measurable random variables such that $\textrm{ess}\sup_{g\in Q_{p(\cdot)}}\{-\alpha_{t}(g)\}=0$;
			\item[(2)] $\varrho^{K}_{t}$ has a robust representation in terms of the minimal function, i.e.
			\begin{equation}\label{58}
				\varrho^{K}_{t}(f)=\textrm{ess}\sup_{g\in Q_{p(\cdot)}}\big\{\langle g,-f\rangle_{t}-\alpha_{t}^{\min}(g)\big\}\quad \textrm{for any}\quad f\in L^{p(\cdot)};
			\end{equation}
			\item[(3)] $\varrho^{K}_{t}$ is continuous from above under $K$, i.e.
			\begin{equation}\label{59}
				f_{n}\searrow f\Rightarrow \varrho^{K}_{t}(f_{n})\nearrow \varrho^{K}_{t}(f).
			\end{equation}
		\end{description}
	}
\end{Theorem}

\noindent \textbf{Proof.}
$(2)\Rightarrow (1)$ is obvious. We first prove $(1)\Rightarrow(3)$. Using Lemma 5 of Cheng and Xu (2013), suppose that $f_{n}\searrow f$. Then, by the monotonicity of $\varrho^{K}_{t}$, we have $\varrho^{K}_{t}(f_{n})\nearrow \varrho^{K}_{t}(f)$.\\
Next, we show $(3)\Rightarrow (2)$. The inequality
\begin{displaymath}
	\varrho^{K}_{t}(f)\geq\textrm{ess}\sup_{g\in Q_{p(\cdot)}}\big\{\langle g,-f\rangle_{t}-\alpha_{t}^{min}(g)\big\}
\end{displaymath}
is a direct consequence of the definition of $\alpha_{t}^{\min}$. Now, we need only show the inverse inequality. To this end, we define a map $\widetilde{\varrho^{K}}: L^{p(\cdot)}\rightarrow \mathbb{R}$ as $\widetilde{\varrho^{K}}(f)=\int_{\Omega}\varrho^{K}_{t}(f)dP$. It is easy to check that $\widetilde{\varrho^{K}}$ is a $p(\cdot)$-convex risk measure as defined in Sect.~\ref{sec:3} which is continuous from above. Hence, by Theorem~\ref{T32}, we know that $\widetilde{\varrho^{K}}$ has the dual representation
\begin{displaymath}
	\widetilde{\varrho^{K}}(f)=\sup_{g\in Q_{p(\cdot)}}\big\{\langle g,-f\rangle-\alpha(g)\big\},\quad f\in L^{p(\cdot)},
\end{displaymath}
where the minimum penalty function $\alpha_{\min}$ is given by $\alpha_{\min}(g):=\sup_{f\in \mathcal{A}_{\widetilde{\varrho^{K}}}}\big\{\langle g,-f\rangle\big\}$. By Lemma~\ref{L51}, we have
\begin{displaymath}
	\int_{\Omega}\alpha_{t}^{\min}(g) dP=\sup_{f\in\mathcal{A}_{t}}\langle g,-f\rangle
\end{displaymath}
for any $g\in Q_{p(\cdot)}$. As $\widetilde{\varrho^{K}}(f)\leq0$ for all $f\in\mathcal{A}_{t}$,
\begin{displaymath}
	\int_{\Omega}\alpha_{t}^{\min}(g) dP=\sup_{f\in\mathcal{A}_{t}}\langle g,-f\rangle\leq\alpha(g)
\end{displaymath}
for any $g\in Q_{p(\cdot)}$. Thus, we have
\begin{eqnarray*}
	\int_{\Omega} \varrho^{K}_{t}(f) dP&=&\widetilde{\varrho^{K}}(f)\\
	&=&\sup_{g\in Q_{p(\cdot)}}\big\{\langle g,-f\rangle-\alpha(g)\big\}\\&\leq&\sup_{g\in Q_{p(\cdot)}}\Big\{\int_{\Omega}\langle g,-f\rangle_{t} dP-\int_{\Omega}\alpha_{t}^{\min}(g)dP\Big\}\\&=&\sup_{g\in Q_{p(\cdot)}}\Big\{\int_{\Omega}\big(\langle g,-f\rangle_{t} -\alpha_{t}^{\min}(g)\big)dP\Big\}\\&\leq&\int_{\Omega}\textrm{ess}\sup_{g\in Q_{p(\cdot)}}\big\{\langle g,-f\rangle_{t}-\alpha_{t}^{\min}(g)\big\}dP.
\end{eqnarray*}
Thus, (\ref{58}) holds.

\begin{Remark}
	The conditional $p(\cdot)$-convex risk measures (with partial order $ \leq_{K} $) are extensions of conditional convex risk measures introduced in Detlefsen and Scandolo (2005).
	In fact, if the position space $ L^{p(\cdot)} $ degenerates into the ordinary $ L^{p} $ space, then the dual representation in Theorem~\ref{T51} will also degenerates into the dual representation of conditional  convex risk measures in Detlefsen and Scandolo (2005).
\end{Remark}

Now, with the definition and dual representation, we consider  the time consistency of dynamic $p(\cdot)$-convex risk measures.

\begin{Definition}\label{D54}\textnormal
	{
		A dynamic $p(\cdot)$-convex risk measure $({\varrho^{K}}_{t})_{t=0}^{T}$ is said to be time consistent if, for all $f_{1}, f_{2}\in L^{p(\cdot)}$ and $0\leq t<t+s\leq T$,
		\begin{equation}\label{510}
			{\varrho^{K}}_{t+s}(f_{1})\leq{\varrho^{K}}_{t+s}(f_{2})\Rightarrow {\varrho^{K}}_{t}(f_{1})\leq{\varrho^{K}}_{t}(f_{2}).
		\end{equation}
	}
\end{Definition}

\begin{Remark}\label{R53}\textnormal
	{
		Time consistency means that if two payoffs will have the same riskiness tomorrow in every state of nature, then the same conclusion should be drawn today.
	}
\end{Remark}

\begin{Theorem}\label{T52}\textnormal
	{
		Let $({\varrho^{K}}_{t})_{t=0}^{T}$ be a dynamic $p(\cdot)$-convex risk measure (with partial order $ \leq_{K} $) such that each ${\varrho^{K}}_{t}$ is continuous from above. Then, the following conditions are equivalent for any $0\leq t<t+s\leq T$:
		\begin{description}
			\item[1).] $(\varrho^{K}_{t})_{t=0}^{T}$ is time consistent;
			\item[2).] $\mathcal{A}_{t}=\mathcal{A}_{t,t+s}+\mathcal{A}_{t+s}$;
			\item[3).] ${\varrho^{K}}_{t}\big(-{\varrho^{K}}_{t+s}(f)z\big)={\varrho^{K}}_{t}(f)$ for any $f\in L^{p(\cdot)}$.
		\end{description}
	}
\end{Theorem}
\noindent \textbf{Proof.}
We first show the equivalence between 1) and 3). Suppose that 3) holds and ${\varrho^{K}}_{t+s}(f_{1})\leq{\varrho^{K}}_{t+s}(f_{2})$ for any $f_{1}, f_{2}\in L^{p(\cdot)}$. Then, by the monotonicity of ${\varrho^{K}}_{t}$,
\begin{displaymath}
	{\varrho^{K}}_{t}(f_{1})={\varrho^{K}}_{t}\big(-{\varrho^{K}}_{t+s}(f_{1})z\big)\leq \varrho^{K}_{t}\big(-{\varrho^{K}}_{t+s}(f_{1})z\big)={\varrho^{K}}_{t}(f_{2}).
\end{displaymath}
Next, suppose that $({\varrho^{K}}_{t})_{t=0}^{T}$ is time consistent, and set $f_{2}:=-{\varrho^{K}}_{t+s}(f_{1})z$  and then by Definition~\ref{D51}, we get
that  $-{\varrho^{K}}_{t+s}(f_{1})z=-{\varrho^{K}}_{t+s}(f_{2})z$ for any $f_{1}\in L^{p(\cdot)}$. Thus,
\begin{displaymath}
	{\varrho^{K}}_{t}(f_{1})={\varrho^{K}}_{t}(f_{2})={\varrho^{K}}_{t}\big(-\varrho^{K}_{t+s}(f_{1})z\big).
\end{displaymath}
We now show the equivalence between 2) and 3). To this end, suppose that 3) holds and let $f_{1}\in \mathcal{A}_{t,t+s}$, $f_{2}\in \mathcal{A}_{t+s}$. Then, setting $f:=f_{1}+f_{2}$, we have
\begin{displaymath}
	{\varrho^{K}}_{t+s}(f)={\varrho^{K}}_{t+s}(f_{1}+f_{2})={\varrho^{K}}_{t+s}(f_{2})-\frac{f_{1}}{z}\leq -\frac{f_{1}}{z} .
\end{displaymath}
Thus, by the monotonicity of $ {\varrho^{K}}_{t}$, we know that
\begin{displaymath}
	{\varrho^{K}}_{t}(f)={\varrho^{K}}_{t}\big(-{\varrho^{K}}_{t+s}(f)z\big)\leq {\varrho^{K}}_{t}(f_{1})\leq0,
\end{displaymath}
which implies
\begin{displaymath}
	\mathcal{A}_{t}\supseteq\mathcal{A}_{t,t+s}+\mathcal{A}_{t+s}.
\end{displaymath}
For the inverse relation, let $f\in \mathcal{A}_{t}$ and define $f_{2}:=f+{\varrho^{K}}_{t+s}(f)z$, $f_{1}:=f-f_{2}=-{\varrho^{K}}_{t+s}(f)z$. Then, by the conditional cash invariance of ${\varrho^{K}}_{t}$, it is easy to check that $f_{1}\in \mathcal{A}_{t,t+s}$, $f_{2}\in \mathcal{A}_{t+s}$, which implies \begin{displaymath}
	\mathcal{A}_{t}\subseteq\mathcal{A}_{t,t+s}+\mathcal{A}_{t+s}.
\end{displaymath}
Let us now suppose that 2) holds and $f\in \mathcal{A}_{t}$. It is easy to check that $f+{\varrho^{K}}_{t+s}(f)z\in \mathcal{A}_{t+s}$. Then, with $\mathcal{A}_{t}\subseteq\mathcal{A}_{t,t+s}+\mathcal{A}_{t+s}$, we have $-{\varrho^{K}}_{t+s}(f)z\in \mathcal{A}_{t,t+s}$. Hence, we know that  ${\varrho^{K}}_{t}\big(-{\varrho^{K}}_{t+s}(f)z\big)\leq0$, which implies
\begin{displaymath}
	{\varrho^{K}}_{t}\big(-{\varrho^{K}}_{t+s}(f)z\big)\leq\varrho^{K}_{t}(f).
\end{displaymath}
Now, we need only show the inverse inequality. Indeed, for any $f\in L^{p(\cdot)}$ such that $-{\varrho^{K}}_{t+s}(f)z\in \mathcal{A}_{t,t+s}$, we have $\varrho^{K}_{t}\big(-\varrho^{K}_{t+s}(f)z\big)\leq0$. It is easy to check that $f+{\varrho^{K}}_{t+s}(f)z\in \mathcal{A}_{t+s}$. Thus, by $\mathcal{A}_{t}\supseteq\mathcal{A}_{t,t+s}+\mathcal{A}_{t+s}$, we have $f\in \mathcal{A}_{t}$, which implies
\begin{displaymath}
	\varrho^{K}_{t}\big(-\varrho^{K}_{t+s}(f)z\big)\geq\varrho^{K}_{t}(f).
\end{displaymath}
\\

The case for the recursive property strongly relies on the validity of conditional cash invariance for $\varrho^{K}_{t}$, and hence on the interpretation as conditional capital requirements. In fact, if $\varrho^{K}_{t+s}(f)$ is the conditional capital requirement that has to be set
aside at date $t+s$ in view of the final payoff $f$, then the risky position is equivalently described, at date $t$, by the payoff $\varrho^{K}_{t}\big(-\varrho^{K}_{t+s}(f)z\big)$ occurring in $t+s$.\\

We end this section with a special example of conditional $p(\cdot)$-convex risk measures (with partial order $ \leq_{K} $).

\begin{Example}\label{E51}\textnormal
	{
		(Conditional OCE) Let $u:E\rightarrow \mathbb{R}$ be a closed, concave, and non-decreasing (partially ordered by $K$) function and suppose that $u(\theta)=0$, where $\theta$ is the zero element of $E$.
		Then, for any $f\in L^{p(\cdot)}$, the conditional OCE of some uncertain outcome $f$ is defined by the map $S_{u}:L^{p(\cdot)}\rightarrow L_{t}^{\infty}$:
		\begin{displaymath}
			S_{u}(f)=\textrm{ess}\sup_{\eta\in L_{t}^{\infty}}\Big\{\eta+\mathbb{E}\big[\big(u(f-\eta z)\big)|\mathcal{F}_{t}\big]\Big\}.
		\end{displaymath}
		Thus, by Definition~\ref{D51}, it is easy to check that the function $\varrho^{K}_{t}$ defined as $\varrho^{K}_{t}(f):=-S_{u}(f)$ for any $f\in L^{p(\cdot)}$ is a conditional $p(\cdot)$-convex risk measure.
	}
\end{Example}


\section{Conclusion}
~~~~~~~In this paper, we introduced risk measures on a unique variable exponent Bochner--Lebesgue space denoted as $L^{p(\cdot)}$ and then explored dynamic and cash sub-additive risk measures in this space. Through further refinement of the axioms associated with these risk measures, we derived their dual representations. Additionally, we delved into the investigation of the optimized certainty equivalent on variable exponent Bochner--Lebesgue spaces as an illustrative example. The study of this paper gives a new set of risk measures to capture the fluctuation of volatility of financial markets and further investigation on how these risk measures can be applied to financial markets can be carried on.

\end{document}